\documentclass[11pt]{article}
\pdfoutput=1
\usepackage{latexsym}
\usepackage{amssymb}
\usepackage{epsf}
\usepackage{amsmath}
\usepackage{graphicx}
\usepackage{slashed}

\newcommand{\lsim}{\lesssim}

\newcommand{\tr}{{\rm Tr}}

\begin{document}
\pagestyle{empty}

\begin{flushright}
KEK-TH-1743
\end{flushright}

\vspace{3cm}

\begin{center}

{\bf\LARGE Partially Composite Dark Matter} 
\\

\vspace*{1.5cm}
{\large 
Masaki Asano$^{1}$ and Ryuichiro Kitano$^{2,3}$
} \\
\vspace*{0.5cm}

{\it
$^1$Physikalisches Institut and Bethe Center for Theoretical Physics\\
Universit\"at Bonn, Nussallee 12, D-53115 Bonn, Germany\\
$^2$KEK Theory Center, Tsukuba 305-0801, Japan\\
$^3$Department of Particle and Nuclear Physics\\
The Graduate University for Advanced Studies (Sokendai)\\
Tsukuba 305-0801, Japan\\
}

\end{center}

\vspace*{1.0cm}

\begin{abstract}
{\normalsize 
In a class of theories where the Higgs field emerges as a pseudo
Nambu-Goldstone boson, it is often assumed that interactions to generate
the top Yukawa coupling provide the Higgs potential as well.
Such a scenario generically requires a little cancellation in the
 leading contribution to the Higgs potential, and the electroweak scale
 is generated by the balance between the leading and the subleading
 contributions.
We, instead, consider the possibility that the contribution from the
 dark matter particle balances against that from the top quark.
The thermal relic of the new particle explains the abundance
 of dark matter in a consistent region of the parameter space, and the
 direct detection is found to be promising.
}
\end{abstract} 

\newpage
\baselineskip=18pt
\setcounter{page}{2}
\pagestyle{plain}
\baselineskip=18pt
\pagestyle{plain}

\setcounter{footnote}{0}

\section{Introduction}

The discovery of the Higgs boson~\cite{Aad:2012tfa,Chatrchyan:2012ufa} 
completes the list of the particles in the Standard Model, but, 
nevertheless, the origin of the Higgs field and its potential remains as 
mysteries. The Higgs boson mass, $m_h = 126$~GeV~\cite{ATLAS:Higgs,CMS:Higgs}, 
which is smaller than its vacuum expectation value (VEV), $v =
246$~GeV, is far from the naive expectation from the analogy of QCD,
$m_h \sim 4 \pi v$, which naturally leads us to consider the possibility
that the Higgs field as a pseudo Nambu-Goldstone boson~\cite{pNGBHiggs}.

The hypothesis of the partially composite fermions~\cite{Kaplan:1991dc}
provides a consistent picture for the scenario of the light Higgs boson
as a pseudo Nambu-Goldstone boson. The dynamically broken global
symmetry to ensure the massless NG boson is explicitly but weakly broken
by couplings between the fermions in the Standard Model and operators in
the dynamical sector in addition to the gauge interactions in the
Standard Model. The coupling induces mixings between the elementary
fermions and composite hadronic states in the dynamical sector, and the
Yukawa interactions as well as the Higgs potential are generated through
the mixing.
The top quark should give the most important contribution to the Higgs
potential since its Yukawa coupling is the largest. It has been studied
that various types of models can indeed reproduce the 126~GeV Higgs
boson while explaining the top quark and the $W$ boson masses. (For a
recent review, see Ref.~\cite{Bellazzini:2014yua}.)

The TeV scale new physics motivated by the origin of the Higgs boson
also provides a natural explanation of dark matter of the Universe as
thermal relic of a stable TeV or weak scale particle.
If there is such a particle in the scenario of the pseudo
Nambu-Goldstone Higgs, one should also consider the contribution to the
Higgs potential from the dark matter particle.
Interestingly, such a contribution is somewhat needed to generate a
realistic potential. The constraints from the electroweak precision
measurements prefer to have a little hierarchy between the scale of
dynamical symmetry breaking, $f$, and the electroweak VEV, $v$, so that
the Standard Model is realized as a good low energy effective theory.
If a single source of the potential is dominated such as from the top
quark, the naive expectation of the Higgs VEV is zero or of the order of
$\pi f$ due to the periodicity of the Higgs potential. Both are clearly
not acceptable.
A little hierarchy can be accommodated by assuming a
cancellation in the leading contribution so that the subleading one
becomes important.
The presence of the dark matter particle provides another
possibility. The Higgs potential is destabilized at the origin by the
contribution from the top quark, and stabilized at a small value $v$ by
that from the dark matter particle.
Dark matter candidates in the models of dynamical electroweak symmetry
breaking have been considered in the literatures; e.g., ``technibaryon"
\cite{techniB1,techniB2} and ``topological dark
matter"~\cite{topologicalDM1,topologicalDM2}\footnote{See also Ref.~\cite{2012uc2014msa}}. 
Also, a Majorana fermion in the strong dynamics as the dark matter particle 
has been discussed in Ref.~\cite{Kouvaris:2007iq}.

In this paper, we study the contributions to the Higgs potential from
the weakly interacting massive particle (WIMP) dark matter.
As a concrete example, we consider the $SO(5)/SO(4)$ model for the
Nambu-Goldstone Higgs field~\cite{Agashe:2004rs}, 
and introduce a gauge singlet Majorana
fermion as the dark matter particle which couples to the strong sector
in a way that $SO(5)$ symmetry is explicitly broken.
The dark matter generates the Higgs potential of the $\sin^2 h/f$ type
at the leading order of the coupling. This contribution can balance
against the $\cos h/f$ type potential generated from the top quark.
We find that in the parameter region where the correct size of the Higgs
potential is generated, the dark matter abundance is explained
simultaneously through the induced coupling between dark matter and the
Higgs field. The predicted spin-independent cross sections for the direct
detection experiments are found to be consistent with the current
experimental bounds, but are large enough to be covered by the future
experiments.

This paper is organized as follows. In the next section, we review the
minimal composite Higgs model~\cite{Agashe:2004rs} which we use for the
basis of our study. In Section 3, we calculate the dark-matter
contribution to the Higgs potential, and discuss the consistent
parameter region in Section 4. The abundance of dark matter and the
possibility of the direct detection are studied in Section 5. Section 6
is devoted to summary and discussion.

\section{The $SO(5)/SO(4)$ model}

We consider the composite Higgs model associated with the $SO(5) \to
SO(4)$ symmetry breaking~\cite{Agashe:2004rs}. 
The unbroken global symmetry $SO(4)$ together with $U(1)_{B-L}$ global
symmetry contains $SU(2)_L \times U(1)_Y$ gauge group as a subgroup.
The unbroken $SO(4)$ symmetry ensures the custodial symmetry in the
strong sector, and thus there is no severe constraints from the
$T$-parameter.

The Nambu-Goldstone field, $\pi (x)$, is introduced as
\begin{align}
 \xi (x)& = e^{i \pi^a (x) X^a}, 
\end{align}
where $X^a$, $a=1,\cdots,4$, are generators of $SO(5)/SO(4)$ in the
vector representation, {\bf 5}, of $SO(5)$~\cite{CCWZ}. 
The $\xi$ field transforms under $SO(5)$ symmetry as
\begin{align}
 \xi & \to \hat g \xi \hat h^{-1} (\pi, \hat g),
\end{align}
where $\hat g \in SO(5)$ and $\hat h \in SO(4)$.  We take the basis where unbroken
$SO(4)$ generators are embedded as
\begin{align}
S^\alpha & = \left(
\begin{array}{cc}
 T^\alpha & 0 \\
 0 & 0 \\
\end{array}
\right),\quad \alpha = 1,\cdots,6.
\end{align}
Therefore the group element $h$ takes the form of
\begin{align}
\hat h & = \left(
\begin{array}{cc}
 * & 0 \\
 0 & 1 \\
\end{array}
\right).
\end{align}

The Higgs field $\Sigma (x)$ is defined as
\begin{align}
 \Sigma (x) & = \xi (x) \left(
\begin{array}{c}
 0\\
 0\\
 0\\
 0\\
 1\\
\end{array}
\right)
 = {\sin (h / f) \over h}\left(
\begin{array}{c}
 h_1 \\
 h_2 \\
 h_3 \\
 h_4 \\
 h \cot (h/ f)\\
\end{array}
\right). 
\end{align}
This field transforms homogeneously as $\Sigma \to \hat g \Sigma$ and
the upper four components have the quantum numbers of the Higgs field in
the Standard Model, and $h^2 = h_1^2 + h_2^2 + h_3^2 + h_4^2$. The
electroweak symmetry breaking is described as $\langle h \rangle =
\langle h_3 \rangle \neq 0$, where $f \sin \langle h / f \rangle =
v = 246$ GeV.

In the original minimal composite Higgs model 
where the top and bottom quarks couple to the operators in
the spinorial representation, {\bf 4}, of $SO(5)$, the Higgs potential
with the following form is generated
\begin{align}
V(h) \simeq \alpha_t \cos {h \over f} - \beta_t \sin^2 {h \over f},
\label{V_top}
\end{align}
where we ignore the small contributions from the SM gauge
interactions. The first and the second terms are the leading and
sub-leading contributions in terms of the expansion with the coupling
constants of the interaction terms between the top quark and the
dynamical sector.  These couplings break the $SO(5)$ symmetry explicitly
since the top and bottom quarks do not fill the complete multiplet of
$SO(5)$. The Higgs potential is generated though the explicit breaking.

By denoting $\lambda_q$ and $\lambda_u$, respectively, as the
dimensionless couplings of $q=(t,b)$ and $t^c$ to the dynamical
sector, the naive estimates of $\alpha_t$ and $\beta_t$ are
\begin{align}
 \alpha_t& = {c_q \lambda_q^2 + c_u \lambda_u^2 \over (4 \pi)^2} N_c m_{t'}^2
 f_{t'}^2,
\quad
 \beta_t = {c_{\beta_t} \lambda_q^2 \lambda_u^2 \over (4 \pi)^2 } 
N_c f_{t'}^4,
\end{align}
where $c_q$, $c_u$ and $c_{\beta_t}$ are unknown $O(1)$ parameters, and
$f_{t'}$ and $m_{t'}$ are coupling and masses of the lowest resonance to
which the operator couples. The top Yukawa coupling is written as
\begin{align}
 y_t& =  {c_t \lambda_q \lambda_u f_{t'}^2  \over m_{t'} f}, 
\end{align}
with an $O(1)$ parameter, $c_t$.

The coefficients $\alpha_t$ and $\beta_t$ are $O(\lambda_{q,u}^2)$ and
$O(\lambda_{q,u}^4)$. On the other hand, from the minimization of the
potential, we find
\begin{align}
 v = 246~{\rm GeV}& = \sqrt{
1 - {\alpha_t^2 \over 4 \beta_t^2} } \times f.
\end{align}
We need $\alpha_t < 2\beta_t$ for the vacuum to be stable, which either
means that the perturbative expansions in terms of $\lambda$'s are
violated or there is some accidental cancellation in $\alpha_t$.
Phenomenologically, one needs $v/f \lsim 0.25$ to satisfy experimental
constraints, especially from the $Zb\bar{b}$
coupling~\cite{Agashe:2005dk}, which means $\alpha_t \simeq 2 \beta_t$.

In the following we consider the possibility that $\alpha_t \gg \beta_t$
as expected from the perturbative expansion, but a large $\sin^2 (h/f)$
term necessary for the electroweak symmetry breaking is supplied by the
contribution from the dark matter particle.

The models in which SM fermions couple to operators in the fundamental,
{\bf 5}, or the antisymmetric, {\bf 10} representation have also been
proposed~\cite{Agashe:2006at,Contino:2006qr} and reported that such
models relax the constraints from the $Zb\bar{b}$ coupling, $v/f <
0.3-0.4$.
For other possibilities see, e.g., Ref.~\cite{Pomarol:2012qf}. In this
paper, we consider the original model in which top and bottom quarks
couple to the operators in the spinorial representation, {\bf 4}, since
that is the simplest option to incorporate the dark matter particle. By
embedding the Standard Model singlet fermion in the {\bf 5}
representation as we explain later, one can generate the potential which
can balance against the contributions from the top quark.
See also Refs.~\cite{Agashe:2004rs,Pomarol:2012qf, Matsedonskyi:2012ym,Redi:2012ha,Marzocca:2012zn,Panico:2012uw,
Pappadopulo:2013vca,Vecchi:2013bja} for
other possible ways for natural electroweak symmetry breaking.

\section{Higgs potential from Dark Matter}

We introduce the dark matter field, $\psi_S$, which is a Majorana
fermion and singlet under the SM gauge group. We assume that the dark
matter field couples to the dynamical sector as
\begin{align}
 {\cal L}& \ni - {m \over 2} \bar \psi_S \psi_S
           +   \lambda \bar \psi_S {\cal O}_5 
           + i \lambda^\prime \bar \psi_S \gamma_5 {\cal O}_5, 
 \label{eq:coupling_DM_DS}
\end{align}
where ${\cal O}_5$ is a Majorana fermionic operator in the dynamical
sector, and is a component of $SO(5)$ vector representation,
\begin{align}
 {\cal O}& = \left(
\begin{array}{c}
 {\cal O}_1 \\
 {\cal O}_2 \\
 {\cal O}_3 \\
 {\cal O}_4 \\
 {\cal O}_5 \\
\end{array}
\right).
\end{align}
The real valued couplings $\lambda$ and $\lambda^\prime$ break the
$SO(5)$ symmetry explicitly. As we will see later, the interaction
between the dark matter and the dynamical sector gives a mass to the
dark matter which we assume to be the dominant contribution. In that
case, one can ignore the mass term $m$ in Eq.~\eqref{eq:coupling_DM_DS},
which in turn makes it possible to eliminate the $\lambda^\prime$ term
by a field redefinition of $\psi_S$.

The 2-point function of $\psi_S$ is written as,
\begin{align}
 \langle \psi_S (x) \bar \psi_S(0) \rangle& =
- \int {d^4 k \over i (2 \pi)^4} 
{e^{-i k x} \over  \slashed k + \lambda^2 \Pi_{55}(k)},
\end{align}
where
\begin{align}
 \Pi_{ij} (q)& = i\int {d^4 x }
\langle {\cal O}_i (x) \bar {\cal O}_j (0) \rangle
{e^{i q x} }
\nonumber \\
& =
\Pi_4 (q) (\delta_{ij} - \Sigma_i \Sigma_j)
+ \Pi_1 (q) \Sigma_i \Sigma_j .
\label{eq:Pi_ij}
\end{align}
In the last expression of Eq.~\eqref{eq:Pi_ij}, we decompose $\Pi$'s in
terms of the unbroken $SO(4)$ symmetry. The field $\Sigma$ is treated as
an external field. The $\Pi_4$ and $\Pi_1$ functions can be expressed in
terms of the spectral functions such as
\begin{align}
 \Pi_4 (q) = & - \int_0^\infty d s
{\slashed q \rho_{4} (s) + \tilde \rho_{4} (s)
+ i \gamma_5 \tilde \rho_{4,5} (s)
\over q^2 - s + i \epsilon}
+ \cdots,
\label{eq:pi4}
\end{align}
\begin{align}
\Pi_1 (q)& = - \int_0^\infty d s
{\slashed q \rho_{1} (s) + \tilde \rho_{1} (s)
+ i \gamma_5 \tilde \rho_{1,5} (s)
\over q^2 - s + i \epsilon}
+ \cdots,
\label{eq:pi1}
\end{align}
where the ellipsis are regular functions of $q^2$, representing the
contact terms\footnote{
The two point functions of Majorana operators satisfy 
\begin{align}
 \Pi(q)& = - C \Pi^{\rm T}(-q) C, 
 \qquad
 \Pi^\dagger (q) = \gamma^0 \Pi(q) \gamma^0, 
\nonumber 
\end{align} 
which forbid the terms proportional to $\gamma_5 \slashed q$. }.

In the case where there is an effective description in terms of
weakly coupled composite states, such as in a large $N$ theory, spectral
functions are approximated as collections of hadron poles:
\begin{align}
& \rho_4 (s) = \sum_i f_{4,i}^2 \delta(s - |m_{4,i}|^2),
\quad
 \rho_1 (s) = \sum_i f_{1,i}^2 \delta(s - |m_{1,i}|^2),
\end{align}
\begin{align}
& \tilde \rho_4 (s) = \sum_i f_{4,i}^2 {\rm Re} [ m_{4,i} ] 
\delta(s - |m_{4,i}|^2), 
\quad
\tilde \rho_1 (s) = \sum_i f_{1,i}^2 {\rm Re} [ m_{1,i} ] 
\delta(s - |m_{1,i}|^2), 
\end{align}
\begin{align}
& \tilde \rho_{4,5} (s) = \sum_i f_{4,i}^2 
{\rm Im} [m_{4,i}] \delta(s - |m_{4,i}|^2),
\quad
\tilde \rho_{1,5} (s) = \sum_i f_{1,i}^2 
{\rm Im} [m_{1,i}] \delta(s - |m_{1,i}|^2).
\end{align}

We assume that $SO(5)$ symmetry is broken by a VEV of some composite
operator, ${X}$, with the mass dimension $d$. Then it contributes to
$\Pi_4 (q) - \Pi_1 (q)$ as $\propto \langle X^\dagger X \rangle / q^{2d
- 1}$ for a large $q$.
This condition gives the Weinberg sum rules for the spectral functions:
\begin{align}
& \int_0^\infty d s (\rho_4 (s) - \rho_1 (s)) = 0, 
&(d>1), 
\label{eq:sum1} \\
& \int_0^\infty d s  (\tilde \rho_4 (s) - \tilde \rho_1 (s)) = 0, \quad 
\int_0^\infty d s  (\tilde \rho_{4,5} (s) - \tilde \rho_{1,5} (s)) = 0,
&(d>3/2), 
\label{eq:sum2} \\
& \int_0^\infty d s \cdot  s (\rho_4 (s) - \rho_1 (s)) = 0, 
&(d>2), 
\label{eq:sum3} \\
& \int_0^\infty d s \cdot s (\tilde \rho_4 (s) - \tilde \rho_1 (s)) = 0,
\quad
 \int_0^\infty d s \cdot s (\tilde \rho_{4,5} (s) - \tilde \rho_{1,5}
 (s)) = 0, 
&(d>5/2).
\label{eq:sum4} 
\end{align}
For example, if $X$ is a fermion pair in an asymptotically free theory,
$d=3$, and the above six sum rules apply.  One can also obtain a
relation that the contact terms in Eqs.~\eqref{eq:pi4} and
\eqref{eq:pi1} are common for $\Pi_4(q)$ and $\Pi_1(q)$.

The Higgs potential can be calculated by using the Coleman-Weinberg formula:
\begin{align}
 V(h)& = - {1 \over 2} \int {d^4 k \over i (2 \pi)^4}
\tr \log [
\slashed k + \lambda^2 \Pi_{55} (k) + i \epsilon
]
\nonumber \\
&={\rm const.} - {1 \over 2} \int {d^4 k \over i (2 \pi)^4}
\tr \left[ {-\lambda^2
\over \slashed k + i \epsilon
} {(\Pi_4(k) - \Pi_1(k)) \Sigma_5 \Sigma_5} \right]
+ O(\lambda^4)
\nonumber \\
& \equiv {\rm const.} - \beta \sin^2 {h \over f} + O(\lambda^4),
\label{V_DM}
\end{align}
where
\begin{align}
 \beta& = 
- {1 \over 2} \cdot \lambda^2 \int_0^\infty d s
\int {d^4 k \over i (2 \pi)^4}
{4 k^2 (\rho_4(s) - \rho_1 (s))
 \over (k^2 + i \epsilon) (k^2 - s + i \epsilon)} .
\end{align}
The Weinberg sum rules make the momentum integral converge. The piece
which is non-vanishing under the Weinberg sum rules~\eqref{eq:sum1} and
\eqref{eq:sum3} is
\begin{align}
  \beta & = 
 {1 \over 2} { 4 \lambda^2 \over (4 \pi)^2}
\int_0^\infty d s \cdot
s (\rho_4 (s) - \rho_1 (s))
\log {s \over s_0},
\end{align}
where $s_0$ is an arbitrary number. The $s_0$ independence is
ensured by Eq.~\eqref{eq:sum3}.

When we set ${s_0}$ as the mass squared of the lowest resonance to
couple the operator ${\cal O}_i$, $m_{\cal O}$,
\begin{align}
\beta &= {1 \over 2}
{ 4 \lambda^2 \over (4 \pi)^2}
\int_{m_{\cal O}^2}^\infty d s \cdot
s (\rho_4 (s) - \rho_1 (s))
\log {s \over m_{\cal O}^2}
\nonumber \\
& = - {1 \over 2} { 4 \lambda^2 \over (4 \pi)^2}
\int_{m_{\cal O}^2}^\infty d s \cdot
\Delta (s) \left(
1 + \log {s \over m_{\cal O}^2}
\right)
,
\label{eq:beta}
\end{align}
where
\begin{align}
 \Delta (s)& = \int_{m_{\cal O}^2}^s d s'
(\rho_4 (s') - \rho_1 (s')).
\end{align}
The function $\Delta (s)$ goes to zero as $s \to \infty$ because of
Eq.~\eqref{eq:sum1}. Therefore,
the integration in Eq.~\eqref{eq:beta} should be dominated by the lower
resonances. Therefore, we expect
\begin{align}
 \beta& = {1 \over 2}{ 4 \lambda^2 \over (4 \pi)^2}
c_\beta m_{\cal O}^2 f_{\cal O}^2,
\label{beta}
\end{align}
where $c_\beta$ is an $O(1)$ coefficient, and $f_{\cal O}$ is the
coupling of the lowest resonance which couples to the operator ${\cal
O}_i$. The overall sign depends on that of $\rho_4(s) - \rho_1(s)$
near $s \sim m_{\cal O}^2$. We assume $\beta > 0$ which is necessary
for the vacuum to be stable.

\section{Electroweak symmetry breaking}

Adding the contribution in Eq.~\eqref{V_DM} to Eq.~\eqref{V_top}, the
total Higgs potential is obtained as
\begin{align}
 V(h)& = \alpha_t \cos {h \over f} - (\beta + \beta_t) \sin^2 {h \over f}.
\end{align}
The minimization of the potential gives the electroweak VEV and the
Higgs mass as follows:
\begin{align}
 v &= 246~{\rm GeV} = \sqrt{
1 - {\alpha_t^2 \over 4 (\beta + \beta_t)^2} } \times f
   \equiv \epsilon f, 
\label{def_vev} \\ 
 m_h^2 & = (126~{\rm GeV})^2 = {2 (\beta + \beta_t) \epsilon^2 \over f^2}.
\end{align}
Therefore, in the case where the dark matter contribution exists, the
stable minimum can be found for $\alpha_t \gg \beta_t$.
A small $\epsilon$ can be obtained when $\beta \simeq \alpha_t \gg
\beta_t$. When $\beta_t$ is negligible, from Eq.~\eqref{beta}, we find
\begin{align}
 m_h^2& = c_\beta \cdot \epsilon^2 \cdot
 {4 \lambda^2 \over (4 \pi)^2} m_{\cal O}^2
\left(
{f_{\cal O} \over f}
\right)^2.
\end{align}
From this, we obtain the mass of the first resonance to be
\begin{align}
 m_{\cal O}& = 4.9~{\rm TeV}
\cdot
c_\beta^{-1/2}
\left(
{\lambda f_{\cal O}
\over 
1~{\rm TeV}
}
\right)^{-1}
\left(
{
\epsilon \over 0.2
}
\right)^{-2}.
\end{align}

On the other hand, from Eq.~\eqref{def_vev}, $\alpha_t$ is required to
satisfy: 
\begin{align}
 \alpha_t& = 2 (\beta + \beta_t) \sqrt{1 - \epsilon^2} 
      \simeq 2 \beta 
           = {m_h^2 v^2 \over \epsilon^4}. 
\end{align}
The mass of the lowest top partner resonance is, therefore, given by
\begin{align}
 m_{t'}& = 2.4~{\rm TeV}
\left(
{c_t \cdot 2 \lambda_q \lambda_u \over c_q \lambda_q^2 + c_u \lambda_u^2}
\right)^{1/3}
\left(
\epsilon \over 0.2
\right)^{-1}
\leq
2.4~{\rm TeV} \left(
{c_t \over \sqrt{ c_q c_u }}
\right)^{1/3}
\left(
\epsilon \over 0.2
\right)^{-1}. 
\end{align}
Here, we have used $m_h = 126$~GeV and $m_t = 173$~GeV.
The correct top quark mass requires
\begin{align}
 {\lambda_q \lambda_u f_{t'}^2
\over 
m_{t'}^2
}& = 0.5 \cdot c_t^{-1}
\left(
\epsilon \over 0.2
\right)^{-1}
\left(
m_{t'} \over 2.4~{\rm TeV}
\right)^{-1}.
\label{eq:correcttop}
\end{align}
The assumption that perturbative expansions by $\lambda$, $\lambda_q$
and $\lambda_u$ make sense requires
\begin{align}
 & {\lambda^2 f_{\cal O}^2 \over m_{\cal O}^2 } < 1, \quad
 {\lambda_{q,u}^2 f_{t'}^2 \over m_{t'}^2 } < 1.
\label{eq:topYukawa}
\end{align}
Compared with Eq.~\eqref{eq:correcttop}, we need somewhat large $c_t$ and/or
$m_{t'}$ for reliable perturbative estimates while explaining the top
quark mass.
We also expect that $m_{\cal O} \sim m_{t'}$ since they are both hadrons
in the same dynamics. 
Putting altogether, we find $\lambda f_{\cal O} \sim 1-2$~TeV and
$m_{\cal O} \sim m_{t'} \sim 2-4$~TeV provides the successful
electroweak symmetry breaking within the perturbative regime.
We will see below that the correct abundance of dark matter is
obtained in the same parameter region.

\section{Dark matter abundance and prospects for direct detection}

By integrating out the dynamical sector, the mass and coupling of the
dark matter particle are generated such as
\begin{align}
 {\cal L}_{\rm eff} = &
- {m_{\rm DM} \over 2} \bar \psi_S \psi_S
- i {m_{\rm DM,5} \over 2} \bar \psi_S \gamma_5 \psi_S
\nonumber \\
&
+ {\kappa \over 2}\bar \psi_S \psi_S \sin^2 {h \over f}
+ {i \kappa_5 \over 2} \bar \psi_S \gamma_5 \psi_S \sin^2 {h \over f}.
\label{eq:L_DM}
\end{align}
At the leading order in $\lambda$ and $\epsilon$, they are given by
\begin{align}
 m_{\rm DM}& = - \lambda^2 \int_0^\infty ds {\tilde \rho_1 (s) \over s},
 \quad
 m_{\rm DM, 5} = - \lambda^2 \int_0^\infty ds {\tilde \rho_{1,5} (s) \over s},
\end{align}
\begin{align}
 \kappa& = \lambda^2 \int_0^\infty ds 
{\tilde \rho_4 (s) - \tilde \rho_1 (s) \over s},
\quad
 \kappa_{5} = \lambda^2 \int_0^\infty ds 
{\tilde \rho_{4,5} (s) - \tilde \rho_{1,5} (s) \over s}.
\end{align}
The couplings to the gauge bosons are suppressed by $\epsilon^2$.
One can eliminate the $m_{\rm DM,5}$ term by the redefinition of
$\psi_S$. In that basis, $m_{\rm DM}$, $\kappa$ and $\kappa_{5}$ are
shifted to
\begin{align}
 m_{\rm DM}& \to \sqrt {m_{\rm DM}^2 + m_{\rm DM,5}^{2}}, \qquad 
      \kappa \to \kappa \cos \tilde{\theta} + \kappa_{5} \sin \tilde{\theta}, \qquad 
  \kappa_{5} \to \kappa \sin \tilde{\theta} + \kappa_{5} \cos \tilde{\theta}, 
\end{align}
where $\tan \tilde{\theta} = m_{\rm DM} / m_{\rm DM, 5}$. In general, $\kappa_5$ cannot be eliminated simultaneously.
Since we expect from the dimensional analysis that the dark matter mass and
couplings are $O(\lambda^2 f_{\cal O}^2 / m_{\cal O})$, we take the
parameters in Eq.~\eqref{eq:L_DM} as
\begin{align}
   m_{\rm DM}& = c_{\rm DM} {\lambda^2 f_{\cal O}^2 \over m_{\cal O}}, \qquad 
 m_{\rm DM, 5} = 0, \qquad 
        \kappa = c_{\kappa} {\lambda^2 f_{\cal O}^2 \over m_{\cal O}}, \qquad 
      \kappa_5 = c_{\kappa_5} {\lambda^2 f_{\cal O}^2 \over m_{\cal O}}, \qquad
\label{eq:DM_param}
\end{align}
with $O(1)$ parameters, $c_{\rm DM}$, $c_{\kappa}$ and
$c_{\kappa_5}$.\footnote{ Phenomenology of dark matter candidates
which have a similar effective interactions has been studied in
Refs.~\cite{LopezHonorez:2012kv,Vecchi:2013xra,deSimone:2014pda,Fedderke:2014wda}. }

\subsection{Relic abundance}

Since we expect $\kappa_5 \sim \kappa$, 
the annihilation via the $\kappa_5$ coupling mainly contributes to
determine the relic density of dark matter because it is an $s$-wave
process whereas the one with $\kappa$ is $p$-wave. The annihilation
cross section is given by
\begin{align}
\langle \sigma_{\rm ann.} {v} \rangle \simeq 
&
   4 s  
 \left ( { \kappa_5 \over f^2 } \right)^2
 { v^2 \over \left( s - m_h^2 \right)^2 + m_h^2 \Gamma_h^2 }  
 { \Gamma_h |_{m_h = \sqrt{s}}  \over \sqrt{s} }
\nonumber \\
&
 + 
 { 1 \over 8 \pi } 
 \left ( { \kappa_5 \over f^2 } \right)^2
 \left(
 1 + { 3 m_h^2 \over s - m_h^2 } 
\right)^2
 \left(
 1 - { 4 m_h^2 \over s } 
\right)^{1/2}, 
\end{align}
for $m_{\rm DM} > m_h$, and $s$ and $\Gamma_h$ are the center of mass
energy ($\sim 2 m_{\rm DM}$) and the total decay width of the Higgs boson, respectively. 
In the limit of heavy dark matter, $m_{\rm DM} \gg m_h$, it simplifies
to
\begin{align}
 &
\langle \sigma_{\rm ann.} {v} \rangle \simeq 
{1 \over 2 \pi}
\left(
{\kappa_5 \over f^2 }
\right)^2.
\end{align}
The dependence on the $m_{\rm DM}$ disappears.

\begin{figure}[t]
 \begin{center}
  \includegraphics[width=9cm]{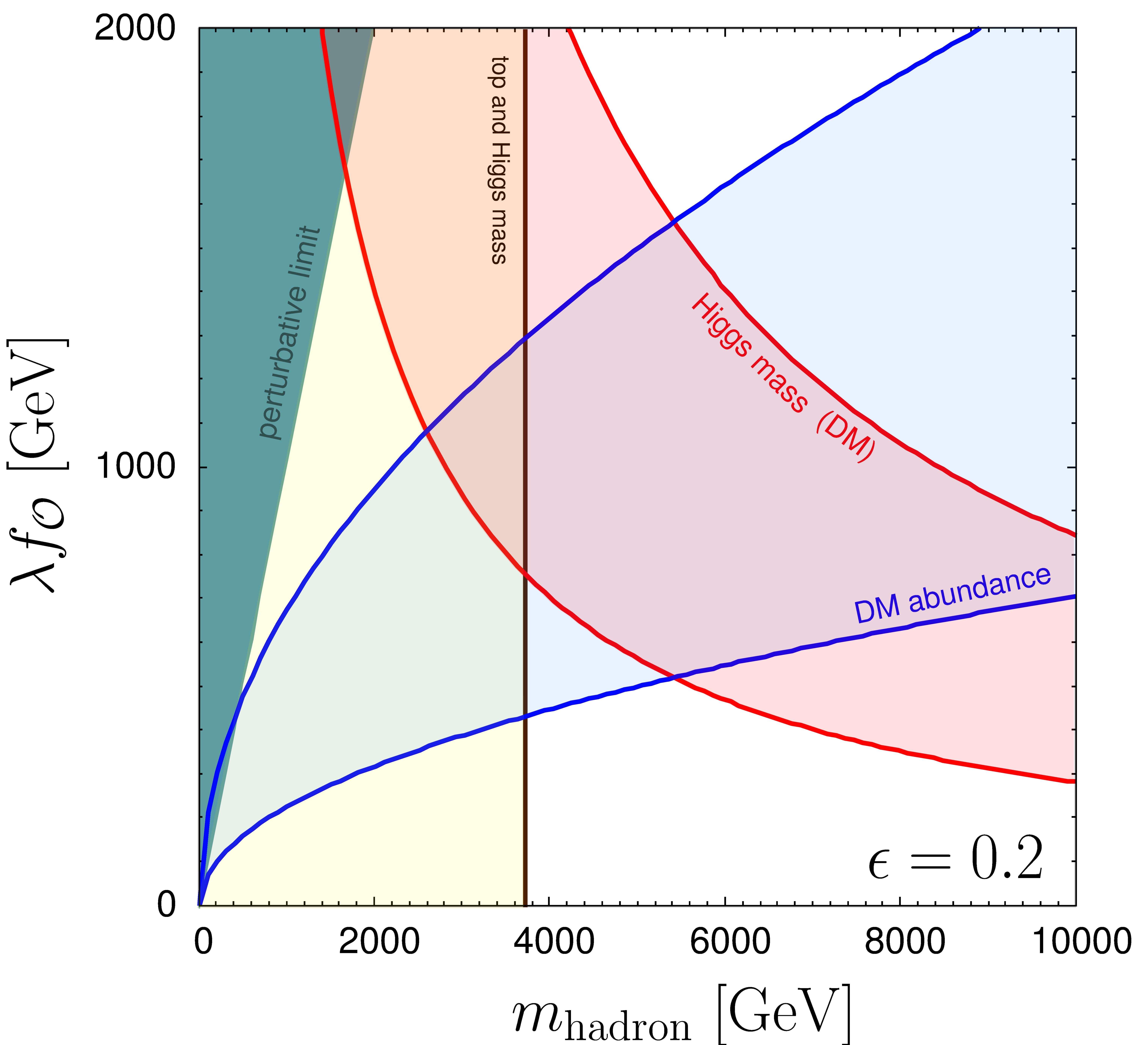}
 \end{center}
\caption{Consistent parameter regions are shown for $\epsilon =
 0.2$. The bands are drawn by taking $O(1)$ parameters to range $1/3 <
 c_X < 3$.}  \label{fig:parameters}
\end{figure}

Requiring 
$\Omega_{{\rm DM}}h^2 = 0.12$~\cite{Ade:2013zuv}, we find
\begin{align}
 \kappa_5& = 160~{\rm GeV} 
\left(
\epsilon \over 0.2
\right)^{-2},
\label{eq:kappa5}
\end{align}
which means
\begin{align}
 {\lambda f_{\cal O}}& = 900~{\rm GeV}
\cdot 
c_{\kappa_5}^{-1/2}
\left(
m_{\cal O} \over 5~{\rm TeV}
\right)^{1/2}
\left(
\epsilon \over 0.2
\right)^{-1}, 
\end{align}
from Eq.~\eqref{eq:DM_param}. Barring $O(1)$ uncertainties in various
estimates, the value is comfortably consistent with the requirements
from the Higgs and top quark masses. We summarize the consistent
parameter regions in Fig.~\ref{fig:parameters} where we allow $O(1)$
parameters to range $1/3 < c < 3$. The requirements from the top quark
mass, the Higgs boson mass, the dark matter abundance all agree in the region
where the perturbative expansion is reliable.

\subsection{Direct detection cross section}

From Eqs.~\eqref{eq:DM_param} and \eqref{eq:kappa5}, the dark matter mass
is obtained as
\begin{align}
m_{\rm DM}& = 160~{\rm GeV} 
\left( 
c_{\rm DM} \over c_{\kappa_5} 
\right)
\left(
\epsilon \over 0.2
\right)^{-2}.
\end{align}
The on-going direct detection experiments have good sensitivity for such
a weak scale dark matter.

In contrast to the case of the annihilation process, the $\kappa$
coupling, rather than $\kappa_5$, mainly contributes to the scattering
processes for the direct detection since the $\kappa_5$ coupling only
contributes to the spin-dependent part in the non-relativistic limit.
The spin independent cross section per nucleon is
given by
\begin{align}
 \sigma_{\rm SI} & = 
 { 4 \over \pi } 
 \left( { m_N m_{\rm DM} \over m_N + m_{\rm DM} } \right)^2
{ \left[ Z f_p + \left( A-Z \right) f_n \right]^2 \over A^2}, 
\nonumber \\ 
f_N & = { \kappa \over f^2 } 
{ m_N \over m_h^2 } 
\left[ 
\sum_{q=u,d,s} f_q^N + \frac{2}{9} \left( 1-\sum_{q=u,d,s}f_q^N \right)
\right], 
\end{align}
where $m_N$ ($N=p,n$) is the nucleon mass and $A$ and $Z$ are the mass 
and atomic number of the target nucleus, respectively. The factor 
$f_q^N$ are matrix elements, 
$f_q^N = (m_q/m_N) \langle N| \bar{q}q |N \rangle$. Assuming 
$f_q^n=f_q^p$ and taking the following values, $f_u^p = 0.021$, 
$f_d^p=0.029$ and $f_s^p=0.009$~\cite{Agrawal:2010fh,Oksuzian:2012rzb}
\footnote{ If we take other values in the literatures, e.g., in Refs.~\cite{Cheng:2012qr}\cite{Belanger:2013oya}, the cross section changes by $O(10\%)$. }, 
which provide us with conservative estimates, the cross section is given by
\begin{align}
 \sigma_{\rm SI} & \simeq 
1.2 \times 10^{-45} {\rm cm^2}
\left(
c_{\kappa} \over c_{\kappa_5}
\right)^{2}, 
\end{align}
where we use the information of the annihilation cross section in 
Eq.~\eqref{eq:kappa5}.

\begin{figure}[htbp]
  \begin{center}
    \includegraphics[width=12cm]{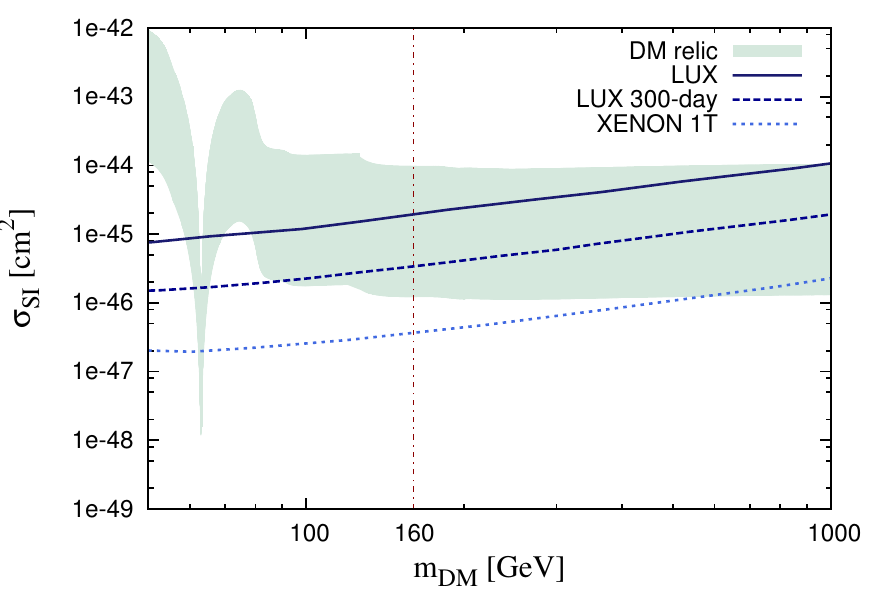}
    \caption{\small Spin independent cross section of the dark matter-nucleon  scattering.}
   \label{fig:DD_sigma}
  \end{center}
\end{figure}

Fig.~\ref{fig:DD_sigma} shows the spin independent cross section of the
dark matter scattering on proton.
In the shaded region, the thermal dark matter abundance is consistent
with the current observation and, in this figure, we assume $ 1/3 <
c_{\kappa}/c_{\kappa_5} < 3$. The solid line and dashed line show the
upper bound on the cross section from the LUX experiment and the
expected upper bound from future LUX $300$-day run,
respectively~\cite{Akerib:2013tjd,Szydagis:2014xog}. The dotted line
denotes the expected upper bound from the future XENON 1T
experiment~\cite{Aprile:2012zx}\footnote{ The constraints from indirect
detections of dark matter turn out to be not quite
strong~\cite{Fedderke:2013pbc}.}.  As one can see, a large parameter
region can be covered by direct detection experiments in near future.

Here, we comment on other possible signatures of the model. The
compositeness of the Higgs boson affects the coupling of the Higgs
boson. In particular, the coupling to electroweak gauge bosons can be
measured with a good accuracy. It has been studied that the sensitivity
can reach to $\epsilon \sim 0.1$ ($0.01$) at the LHC
(ILC)~\cite{Bellazzini:2014yua}\footnote{ For details, see also
Ref.~\cite{Dawson:2013bba}.  }. In the $m_{\rm DM} < m_h/2$ region, the
invisible branching ratio of the Higgs boson decay is also expected. The
current searches at the
LHC~\cite{CMS:H_inv_1,CMS:H_inv_2,Aad:2014iia,ATLAS:H_inv} put an upper
bound around $20 \%$~\cite{Belanger:2013xza} from the global fit
assuming that the coupling constants are not modified from the Standard
Model ones.

Multi-TeV spin-1/2 resonances are expected in the top quark and dark
matter sector as discussed above. The top partner with the same charge
of top quark may be accessible at future collider experiments ($< 3.2$
TeV by LHC 33 TeV with 3
ab$^{-1}$~\cite{Bhattacharya:2013iea,Agashe:2013hma}).

\section{Summary}

Dark matter of the Universe and the Higgs boson are two mysterious items
in particle physics, and probably hints for deeper understanding of
particle physics are hidden there. Indeed, the size of the interaction
required to explain the abundance of dark matter by the thermal relic is
of the order of the weak interaction, that is characterized by the Higgs
VEV. This may be telling us that the nature of dark matter and that of
the Higgs boson are tightly related.

We consider the possibility that dark matter is the one which is
responsible for creating the potential of the Higgs field. We see that
in the minimal composite Higgs model, the balance between the potentials
made by the top quark and the dark matter can trigger the successful
electroweak symmetry breaking while explaining the abundance of the dark
matter.

We have studied the effective theory where the Higgs field is described
as the pseudo Nambu-Goldstone boson. In the language of the effective
theory, the Higgs field is already introduced as effective degrees of
freedom, and we assume that the dark matter particle couples to it
through some interaction term to break the global symmetry.  In a full
dynamical description, however, the picture may be more dramatic; the
Higgs field may actually be the condensation of dark matter. For
example, it has been studied recently that the scenario of the top quark
condensation can have a picture of the pseudo Nambu-Goldstone Higgs
boson~\cite{Cheng:2013qwa, Fukano:2013aea}. It is promising that the
realization of the dark-matter condensation as the Higgs field is also
possible.

In the parameter region where the Higgs boson mass and the abundance of
the dark matter is explained, the spin-independent cross section for the
direct detection experiments turns out to be quite large, just below the
experimental constraints. If there is no significant fine-tuning, we
expect to see the detection quite soon.

\section*{Acknowledgments}

This work is supported by JSPS Grant-in-Aid for Young Scientists (B)
(No. 23740165 [RK]), MEXT Grant-in-Aid for Scientific
Research on Innovative Areas (No. 25105011 [RK]) and the 
German Research Foundation through TRR33 "The Dark Universe"
(MA).


\end{document}